\documentclass[twocolumn,floats,showpacs,superscriptaddress,pre]{revtex4}
\usepackage{amsmath,amssymb,bm}
\usepackage{graphics}
\usepackage{epsfig}
\usepackage{graphicx}

\begin{document}

\title{Quantum Principle of Least Action in Dynamic Theories With Higher
Derivatives}
\author{Natalia Gorobey}
\affiliation{Peter the Great Saint Petersburg Polytechnic University, Polytekhnicheskaya
29, 195251, St. Petersburg, Russia}
\author{Alexander Lukyanenko}
\email{alex.lukyan@mail.ru}
\affiliation{Peter the Great Saint Petersburg Polytechnic University, Polytekhnicheskaya
29, 195251, St. Petersburg, Russia}
\author{A. V. Goltsev}
\affiliation{Ioffe Physical- Technical Institute, Polytekhnicheskaya 26, 195251, St.
Petersburg, Russia}

\begin{abstract}
A generalized canonical form of action of dynamic theories with higher
derivatives is proposed, which does not require the introduction of
additional dynamic variables. This form is the initial point for the
construction of quantum theory, in which the state of motion of the system
is described by the wave functional on its trajectories in the configuration
space, and the functional itself is an eigenvector of the action operator.
The Pais-Uhlenbeck oscillator is considered as the simplest model. For this
case, the correspondence between the new form of quantum theory and
\textquotedblleft ordinary\textquotedblright\ quantum mechanics has been
established in the local limit, when the Pais-Uhlenbeck oscillator is
reduced to the \textquotedblleft ordinary\textquotedblright\ harmonic
oscillator.
\end{abstract}

\maketitle

%\author{Natalia Gorobey$^{1}$, Alexander Lukyanenko$^{1,\ast }$, and A.V.
%Goltsev$^{2}$}
%\email{$^{\ast }$alex.lukyan@mail.ru}
%\affiliation{$^{1}$Peter the Great Saint Petersburg Polytechnic University,
%Polytekhnicheskaya 29, 195251, St. Petersburg, Russia\\
%$^{2}$Ioffe Physical-Technical Institute, Polytekhnicheskaya 26, 195251, St.
%Petersburg, Russia}

%\date{\today }

%\pacs{}

%\begin{multicols}{2}
%\narrowtext

%%%%%%%%%%%%%%%%%%%%%%%%%%%%%%%%%%%%%%%%%%%%%%%%%%%%%%%%%%%%%%%%%%%%%%%

%%\bigskip

\section{\textbf{INTRODUCTION}}

The canonical quantization of classical dynamic theory is based on the
canonical (Hamiltonian) form of its dynamics, in which the dynamic variables
are the canonical coordinates and momenta $\left( q_{k},p_{k}\right) $, and
the dynamics are determined by the Hamiltonian function $h\left( q,p\right) $%
. Here we do not consider singular theories in which connections are
initially present \cite{Dirac}, although this does not exclude their
appearance at some stages. The starting point for quantization is the action
in canonical form:

\begin{equation}
I\left[ q,p\right] =\int_{0}^{T}dt\left[ \overset{\cdot }{q}%
_{k}p_{k}-h\left( q,p\right) \right] .  \label{1}
\end{equation}%
However, there is a large class of important dynamic theories \cite{Stel},
the reduction of which to the standard canonical form Eq.(\ref{1}) is
associated with certain difficulties. These, in particular, include theories
with higher (above the first) derivatives of dynamic variables with respect
to time in the original Lagrange function; $L\left( q,\overset{\cdot }{q},%
\overset{\cdot \cdot }{q},...\right) $. In this case, the difficulty was
overcome by Ostrogradsky \cite{Ostro} by introducing additional variables
and corresponding constraints:

\begin{equation}
\overset{\cdot }{y}=\overset{\cdot \cdot }{q},\overset{\cdot }{y}_{1}=%
\overset{\cdot \cdot \cdot }{q},...,\overset{\cdot }{q}=y,\overset{\cdot }{y}%
=y_{1},...  \label{2}
\end{equation}%
This approach was used in \cite{BuhLyah} when quantizing gravity, taking
into account contributions quadratic in curvature, which contain
second-order derivatives of the metric tensor in time.

By considering higher derivative theories in the broader context of nonlocal
theories, it is possible to propose an alternative form of canonical
representation of action and corresponding quantization. In \cite{GLG}, the
quantum principle of least action (QPLA) was formulated, which in the case
of a local theory with the canonical form of the action Eq.~(\ref{1}) is equivalent
to \textquotedblleft ordinary\textquotedblright\ quantum mechanics in the
Schr\"{o}dinger formulation. In the case of a nonlocal theory, the QPLA
serves as a replacement for the Schr\"{o}dinger equation. This paper
proposes an alternative canonical representation and quantization for the
simplest mechanical theory with higher derivatives -- the Pais-Uhlenbeck
(PU) oscillator \cite{PaUl}.

The next section discusses Ostrogradsky's method as applied to the
Pais-Uhlenbeck oscillator. In the second section, the quantum principle of
least action is formulated for the PU oscillator. In the third section, a
system of equations is obtained that determines the basic state of motion of
the PU oscillator. In the fourth section, the correspondence of the QPLA
with the quantum theory of a harmonic oscillator in the local limit is
established.

\section{OSTROGRADSKY METHOD FOR THE PAIS-UHLENBECK OSCILLATOR}

The Lagrange function of the PU oscillator has the form:

\begin{equation}
L=\frac{1}{2}\left( \overset{\cdot }{q}^{2}-r^{2}\overset{\cdot \cdot }{q}%
^{2}-q^{2}\right) ,  \label{3}
\end{equation}%
where we left the only parameter $r$, which has the dimension of time. When $%
r=0$, we obtain the Lagrange function of the harmonic oscillator. Following
Ostrogradsky's method, we introduce an additional variable $y=\overset{\cdot
}{q}$ and modify the Lagrange function with the corresponding additional
condition:

\begin{equation}
\widetilde{L}=\frac{1}{2}\left( \overset{\cdot }{q}^{2}-r^{2}\overset{\cdot }%
{y}^{2}-q^{2}\right) +\lambda \left( y-\overset{\cdot }{q}\right) ,
\label{4}
\end{equation}%
where $\lambda $ is the indeterminate Lagrange multiplier. Now let us carry
out the standard canonical analysis \cite{Dirac} of the modified theory. It
begins with the definition of canonical momenta:

\begin{equation}
p_{q}=\overset{\cdot }{q}-\lambda ,p_{y}=-r^{2}\overset{\cdot }{y}%
,p_{\lambda }=0.  \label{5}
\end{equation}%
The last equation in the Eq.(\ref{5}) is the primary constraint according to
the classification \cite{Dirac}. We find the Hamilton function using the
Legendre transformation:

\begin{eqnarray}
\widetilde{h} &=&\overset{\cdot }{q}p_{q}+\overset{\cdot }{y}p_{y}-%
\widetilde{L}  \notag \\
&=&\frac{1}{2}\left( p_{q}+\lambda \right) ^{2}-\frac{1}{2r^{2}}%
p_{y}^{2}-\lambda y+\frac{1}{2}q^{2}.  \label{6}
\end{eqnarray}%
The primary constraint must be maintained over time:

\begin{equation}
\left\{ p_{\lambda },\widetilde{h}\right\} =y-\left( p_{q}+\lambda \right)
=0,  \label{7}
\end{equation}%
where curly braces indicate Poisson brackets. As a result, we obtain a
secondary constraint, which, together with the primary constraint (both
constraints of the second kind according to the classification \cite{Dirac},
allows us to exclude the canonical pair $\left( \lambda ,p_{\lambda }\right)
$ from the dynamics. Finally, the Hamilton function of the modified theory
takes the form:

\begin{equation}
\widetilde{h}\left( q,y,p_{q},p_{y}\right) =-\frac{1}{2}\left( \frac{%
p_{y}^{2}}{r^{2}}+y^{2}\right) +\frac{1}{2}q^{2}+yp_{q}.  \label{8}
\end{equation}%
It can be seen that the energy of the PU oscillator is not limited from
below. However, one can apply the standard rules of canonical quantization
\cite{B} and write the Schr\"{o}dinger equation for the wave function of the
system $\psi \left( t,q,y\right) $, described by two dynamic variables.
Increasing the number of arguments of the wave function complicates the
interpretation of quantum theory. Here, an equivalent formulation of quantum
theory is also possible in terms of the wave functional on trajectories,
which requires an expansion of the configuration space: $\Psi \left[ q\left(
t\right) ,y\left( t\right) \right] $. For a local theory with the Hamilton
function Eq.(\ref{8}), this functional is multiplicative \cite{GLG}:

\begin{equation}
\Psi \left[ q\left( t\right) ,y\left( t\right) \right] =\prod\limits_{t}\psi
\left( t,q\left( t\right) ,y\left( t\right) \right) .  \label{9}
\end{equation}%
In the local limit $r\rightarrow 0$, the canonical form, like the original
one, should be reduced to the theory of a harmonic oscillator. We can see
this in the transformation of the classical Hamilton function Eq.(\ref{8}).
In order for this value to remain finite in the limit, $p_{y}=0$, should be
set as an additional condition. Maintaining this condition over time gives a
secondary constraint:

\begin{equation}
\left\{ p_{y}\widetilde{,h}\right\} =y-p_{q}=0.  \label{10}
\end{equation}%
Both of these constraints exclude additional canonical variables $y,p_{y}$
from the dynamics, as a result of which the Hamilton function Eq.~(\ref{8})
has as a limit the Hamilton function of the harmonic oscillator

\begin{equation}
\frac{1}{2}\left( p_{q}^{2}+q^{2}\right) .  \label{11}
\end{equation}

\section{QUANTUM PRINCIPLE OF LEAST ACTION}

Let's return to the original theory with the Lagrange function Eq.~(\ref{3})
and consider another version of its canonical representation, which does not
require the introduction of additional variables. As the main initial
object, we take the classical action as a functional of trajectory, 
$I\left[ q\left( t\right) \right] $. 
%These quantities are fixed at the ends of the considered
%time interval $t\in \lbrack 0,T]$ when finding the extremum of the action.
In contrast to the usual approach, we define the generalized canonical
momentum as the variational derivative of the action with respect to
velocity,

\begin{equation}
p\left( t\right) \equiv \frac{\delta I}{\delta \overset{\cdot }{q}\left(
t\right) }=\overset{\cdot }{q}\left( t\right) +\omega ^{2}\overset{\cdot
\cdot \cdot }{q}\left( t\right) \equiv \mathit{L}\overset{\cdot }{q}\left(
t\right) .  \label{12}
\end{equation}%
We also need to express the velocity in terms of momentum, which is easily done
by solving the differential equation Eq.(\ref{12}) with given boundary
conditions $\overset{\cdot }{q}\left( 0\right) =v_{0},\overset{\cdot }{q}%
\left( T\right) =v_{T}$:

\begin{eqnarray}
\overset{\cdot }{q}\left( t\right) &=&\mathit{L}^{-1}p\left( t\right) =\frac{%
\sin \omega t}{\sin \omega T}\left( v_{T}-v_{0}\cos \omega T\right)
+v_{0}\cos \omega T  \notag \\
&&+\int_{0}^{T}\mathit{K}\left( t,s\right) p\left( s\right) ds,  \label{13}
\end{eqnarray}%
where

\begin{eqnarray}
\mathit{K}\left( t,s\right) &=&\frac{\omega }{\sin (\omega T)}\left[ \theta
\left( s-t\right) \sin (\omega t) \sin [\omega \left( s-T\right) \right].  \notag
\\
&&\left. +\theta \left( t-s\right) \sin (\omega s)\sin (\omega \left(
t-T\right)) \right] ,  \label{14}
\end{eqnarray}%
$\omega =r^{-1}$, and

\begin{equation}
\theta \left( x\right) =\left\{
\begin{array}{c}
1,x>0 \\
0,x<0%
\end{array}%
\right. .  \label{15}
\end{equation}%
Further, to simplify the consideration, let us set $v_{0}=v_{T}=0$ (the
return points of the oscillator oscillations), so that $\mathit{L}^{-1}$
will be a linear operator. Let us now introduce the Hamilton functional
through the generalized Legendre transformation:

\begin{equation}
\mathit{H}\left[ q\left( t\right) ,p\left( t\right) \right] =\int_{0}^{T}dt%
\overset{\cdot }{q}\left( t\right) p\left( t\right) -I\left[ q\left(
t\right) ,\overset{\cdot }{q}\left( t\right) \right] ,  \label{16}
\end{equation}%
where the velocity $\overset{\cdot }{q}\left( t\right) $ in both terms on the
right is considered to be expressed in terms of momentum according to the
Eq.(\ref{13}). As a result we find:

\begin{eqnarray}
\mathit{H}\left[ q\left( t\right) ,p\left( t\right) \right] &=&\frac{1}{2}%
\left[ \int_{0}^{T}dt\int_{0}^{T}dsp\left( t\right) \mathit{K}\left(
t,s\right) p\left( s\right) \right.  \notag \\
&&\left. +q\left( t\right) \delta \left( t-s\right) q\left( s\right) \right]
.  \label{17}
\end{eqnarray}%
The generalized canonical form of the action is now defined as follows:

\begin{equation}
I\left[ q\left( t\right) ,p\left( t\right) \right] =\int_{0}^{T}dt\overset{%
\cdot }{q}\left( t\right) p\left( t\right) -\mathit{H}\left[ q\left(
t\right) ,p\left( t\right) \right] .  \label{18}
\end{equation}%
In this form of action, $\overset{\cdot }{q}\left( t\right) $ is not an
independent dynamic variable.

The transition to quantum theory is carried out by replacing the classical
functions $q(t)$ and $p(t)$ by operators \cite{GLG}

\begin{equation}
\widehat{q}\left( t\right) \equiv q\left( t\right) ,\widehat{p}\left(
t\right) =\frac{\widetilde{\hslash }}{i}\frac{\delta }{\delta q\left(
t\right) },  \label{19}
\end{equation}%
which act on the space of wave functionals $\Psi \left[ q\left( t\right) %
\right] $. The first of them is the operator of multiplication by the
coordinate $q(t)$, and the second is the operator of variational
differentiation. This operator is defined by the following representation of
the variation of the wave functional:

\begin{equation}
\delta \Psi =\int_{0}^{T}dt\frac{\delta \Psi }{\delta q\left( t\right) }%
\delta q\left( t\right) .  \label{20}
\end{equation}%
From dimensional considerations it follows that the constant $\widetilde{%
\hslash }$ in the Eq.(19) differs from the \textquotedblleft
ordinary\textquotedblright\ Planck constant by a factor having the dimension
of time (see also \cite{FeyHib}):

\begin{equation}
\widetilde{\hslash }=\alpha r\hslash .  \label{21}
\end{equation}%
Here we use the dimensional parameter $r$, which is at our disposal. The
dimensionless parameter $\alpha $ serves to harmonize a new form of nonlocal
quantum theory with the local quantum theory of a harmonic oscillator \cite%
{LanLif}, which is obtained in the limit $r\rightarrow 0$. This
correspondence is established below.

The quantum principle of least action is formulated as a secular equation
for the action operator. This operator is obtained by replacing the dynamic
variables $q(t)$ and $p(t)$ in the action functional Eq.(\ref{16}) with the operators
Eq.(\ref{19}):

\begin{equation}
\widehat{I}\equiv \int_{0}^{T}dt\overset{\cdot }{q}\left( t\right) \frac{%
\widetilde{\hslash }}{i}\frac{\delta }{\delta q\left( t\right) }-\mathit{H}%
\left[ q\left( t\right) ,\frac{\widetilde{\hslash }}{i}\frac{\delta }{\delta
q\left( t\right) }\right] .  \label{22}
\end{equation}%
Thus, the wave functional $\Psi \left[ q\left( t\right) \right] $, which
determines the state of motion of the PU oscillator in this quantum theory,
is found as an eigenvector of the action operator from the equation

\begin{equation}
\widehat{I}\Psi =\Lambda \Psi .  \label{23}
\end{equation}%
The corresponding eigenvalue $\Lambda $, by definition, is a quantity
independent of the interior points of the trajectory $q(t)$. $\Lambda $
depends only on the boundary points $q_{0},q_{T}$ (recall that $%
v_{0}=v_{T}=0 $).

The interpretation of the wave functional $\Psi \left[ q\left( t\right) %
\right] $ is similar to the interpretation of the wave function: if the
normalization condition for the probability is satisfied,

\begin{equation}
\left\vert \left\vert \Psi \right\vert \right\vert ^{2}\equiv \int
\prod\limits_{t}dq\left( t\right) \left\vert \Psi \left[ q\left( t\right) %
\right] \right\vert ^{2}=1.  \label{24}
\end{equation}%
Then $\left\vert \Psi \left[ q\left( t\right) \right] \right\vert ^{2}$ is
the probability density that the oscillator moves in a small neighborhood of
the trajectory $q(t)$.

\section{GROUND STATE OF MOTION OF THE PAIS-UHLENBECK OSCILLATOR}

The ground state of motion of a harmonic oscillator, corresponding to the
minimum energy $\hslash /2$, is described by the wave function \cite{LanLif}

\begin{equation}
\psi _{0}\left( t,q\right) =A\exp \left( -i\frac{t}{2}-\frac{q^{2}}{2\hslash
}\right) .  \label{25}
\end{equation}%
By analogy, we look for a solution to the secular equation Eq.(\ref{23}) for
the wave functional in the form:

\begin{eqnarray}
\Psi _{0}\left[ q\left( t\right) \right] &=&A\exp \left[ -\frac{1}{2%
\widetilde{\hslash }}\int_{0}^{T}dt\int_{0}^{T}dsq\left( t\right) \text{%
\textit{M}}\left( t,s\right) q\left( s\right) \right.  \notag \\
&&\left. +\frac{i}{\widetilde{\hslash }}\int_{0}^{T}dt\mathit{k}\left(
t\right) q\left( t\right) \right] .  \label{26}
\end{eqnarray}%
The action of the kinetic energy operator (the quadratic form of the momenta
in Eq.(\ref{15})) will give a factor in front of the exponent equal to the
sum

\begin{eqnarray}
&&\frac{1}{2}\int_{0}^{T}dt\int_{0}^{T}ds\int_{0}^{T}du\int_{0}^{T}dv  \notag
\\
&&\times q\left( v\right) \text{\textit{M}}\left( v,t\right) \mathit{K}%
\left( t,s\right) \text{\textit{M}}\left( s,u\right) q\left( u\right)  \notag
\\
&&-i\int_{0}^{T}dt\int_{0}^{T}ds\int_{0}^{T}du\mathit{k}\left( t\right)
\mathit{K}\left( t,s\right) \text{\textit{M}}\left( s,u\right) q\left(
u\right)  \notag \\
&&-\frac{1}{2}\int_{0}^{T}dt\int_{0}^{T}ds\mathit{K}\left( t,s\right) \left[
\widetilde{\hslash }\text{\textit{M}}\left( s,t\right) +\mathit{k}\left(
t\right) \mathit{k}\left( s\right) \right] .  \label{27}
\end{eqnarray}%
The action of the first term in the Eq.(\ref{16}) will add to this the sum
of two more terms:

\begin{eqnarray}
&&i\int_{0}^{T}dt\int_{0}^{T}du\overset{\cdot }{q}\left( t\right) \text{%
\textit{M}}\left( t,u\right) q\left( u\right) +\int_{0}^{T}dt\overset{\cdot }%
{q}\left( t\right) \mathit{k}\left( t\right)  \notag \\
&=&-\frac{i}{2}\int_{0}^{T}dt\int_{0}^{T}duq\left( t\right) \left[ \frac{%
\partial }{\partial t}\text{\textit{M}}\left( t,u\right) +\frac{\partial }{%
\partial u}\text{\textit{M}}\left( t,u\right) \right] q\left( u\right)
\notag \\
&&+i\int_{0}^{T}du\left[ q\left( t\right) \text{\textit{M}}\left( t,u\right) %
\right] \left\vert
\begin{array}{c}
t=T \\
t=0%
\end{array}%
\right. q\left( u\right)  \notag \\
&&-\int_{0}^{T}dtq\left( t\right) \overset{\cdot }{\mathit{k}}\left(
t\right) +\left[ q\left( t\right) \mathit{k}\left( t\right) \right]
\left\vert
\begin{array}{c}
t=T \\
t=0%
\end{array}%
\right. .  \label{28}
\end{eqnarray}%
Putting everything together (including the potential energy functional in
the Eq.(\ref{15})), and equating the coefficients at the same powers of $%
q(t) $, we obtain a system of equations:

\begin{eqnarray}
&&\frac{1}{2}\int_{0}^{T}dt\int_{0}^{T}ds\text{\textit{M}}\left( u,t\right)
\mathit{K}\left( t,s\right) \text{\textit{M}}\left( s,v\right)  \notag \\
&=&\frac{i}{2}\left[ \frac{\partial }{\partial u}\text{\textit{M}}\left(
u,v\right) +\frac{\partial }{\partial v}\text{\textit{M}}\left( u,v\right) %
\right] +\frac{1}{2}\delta \left( u-v\right) ,  \label{29}
\end{eqnarray}

\begin{eqnarray}
&&-i\int_{0}^{T}dt\int_{0}^{T}ds\mathit{k}\left( t\right) \mathit{K}\left(
t,s\right) \text{\textit{M}}\left( s,u\right)  \notag \\
&&+i\left[ q\left( t\right) \text{\textit{M}}\left( t,u\right) \right]
\left\vert
\begin{array}{c}
t=T \\
t=0%
\end{array}%
\right. -\overset{\cdot }{\mathit{k}}\left( t\right)  \notag \\
&=&0,  \label{30}
\end{eqnarray}

\begin{eqnarray}
\Lambda &=&-\frac{1}{2}\int_{0}^{T}dt\int_{0}^{T}ds\mathit{K}\left(
t,s\right) \left[ \widetilde{\hslash }\text{\textit{M}}\left( s,t\right) +%
\mathit{k}\left( t\right) \mathit{k}\left( s\right) \right]  \notag \\
&&+\left[ q\left( t\right) \mathit{k}\left( t\right) \right] \left\vert
\begin{array}{c}
t=T \\
t=0%
\end{array}%
\right. .  \label{31}
\end{eqnarray}%
These equations are sufficient to determine the ground state $\Psi _{0}\left[
q\left( t\right) \right] $ and the corresponding eigenvalue $\Lambda $.

\section{LOCAL LIMIT OF THE GROUND STATE OF THE PAIS-UHLENBECK OSCILLATOR}

We are interested, first of all, in the transition to the \textquotedblleft
ordinary\textquotedblright\ quantum theory of a harmonic oscillator in the
limit $r\rightarrow 0$. Obviously, for $r=0$ there should be $\mathit{K}%
\left( t,s\right) =\delta \left( t-s\right) $. In this case, \textit{M}$%
\left( s,t\right) =\delta \left( s-t\right) $ satisfies equation Eq.(\ref{29}%
). Equations Eqs.(\ref{30}) and (\ref{31}) require more careful
consideration. In the specified limit, the first term on the left-hand side of
the Eq.(\ref{28}) does not require integration by parts and is equal to

\begin{equation}
i\frac{q^{2}}{2}\left\vert
\begin{array}{c}
t=T \\
t=0%
\end{array}%
\right.  \label{32}
\end{equation}%
The quantity Eq.(\ref{32}) should be added to the right-hand side of the Eq.(\ref%
{31}) as part of $\Lambda $. For this reason, the second term on the left-hand
side of the Eq.(\ref{30}) is also missing. But then this equation is
satisfied by $\mathit{k}\left( t\right) =0$.

Let's find the limiting value of $\Lambda $. To do this, let us consider in
more detail the limit of the first term in the Eq.(\ref{31}). From equation
Eq.(\ref{29}) we can conclude that in the limit $r\rightarrow 0$ (when the
first term on its right-hand side is equal to zero), the kernel \textit{M}$%
\left( s,t\right) $ of the quadratic form of the wave functional corresponds
to the operator $\sqrt{\mathit{L}}$ (since $\mathit{K}\left( t,s\right) $
corresponds operator $\mathit{L}^{-1}$ ). Thus, as a result, this integral
(recall that $\mathit{k}\left( t\right) =0$) is equal to the trace of the
operator $\sqrt{\mathit{L}^{-1}}$. In order to calculate this trace, we find
the eigenvalues $\lambda _{n}$ of the operator $\mathit{L}$ on the interval $%
[0,T]$ from the equation

\begin{equation}
\overset{\cdot \cdot }{v}+\omega ^{2}v=\omega ^{2}\lambda v.  \label{33}
\end{equation}%
Taking into account the boundary conditions $v_{0}=v_{T}=0$ accepted at the
very beginning, we find the eigenfunctions and the corresponding eigenvalues:

\begin{equation}
v_{n}=A_{n}\sin \frac{\pi nt}{T},\lambda _{n}=1-n^{2}\pi ^{2}\left( \frac{r}{%
T}\right) ^{2}.  \label{34}
\end{equation}%
The trace of the operator $\sqrt{\mathit{L}^{-1}}$ is found as the sum of
the series

\begin{equation}
\sum\limits_{n}\frac{1}{\sqrt{1-n^{2}\pi ^{2}\left( \frac{r}{T}\right) ^{2}}}%
.  \label{35}
\end{equation}%
For small but finite values of $r$, only the first terms of the series are
real, so the required sum is a complex number. However, as $r\rightarrow 0$
the imaginary part tends to zero. Let us approximately calculate the real
part by replacing sum Eq.(\ref{35}) with the integral

\begin{equation}
\frac{\pi T}{r}\int_{-1}^{1}\frac{dx}{\sqrt{1-x^{2}}}=\frac{\pi ^{2}T}{r}.
\label{36}
\end{equation}%
This result is the more accurate the smaller $r$. Then, taking into account the
Eq.(\ref{21}), we obtain the limit of the first term in the Eq.(\ref{31}):

\begin{equation}
\frac{\hslash }{2}\alpha \pi ^{2}T.  \label{37}
\end{equation}%
Now we fix the constant $\alpha $ from the condition that this limit
corresponds to the phase of the wave function of the harmonic oscillator in
\textquotedblleft ordinary\textquotedblright\ quantum mechanics. Let us use
the expression given in \cite{GLG} for the eigenvalue of the action operator
in this case,

\begin{equation}
\Lambda =\frac{\hslash }{i}\left[ \ln \psi \left( T\right) -\ln \psi \left(
0\right) \right] .  \label{38}
\end{equation}%
It is easy to see that the eigenvalue we obtained in the limit $r\rightarrow
0$

\begin{equation}
\Lambda =\frac{\hslash }{2}\alpha \pi ^{2}T+i\frac{q^{2}}{2}\left\vert
\begin{array}{c}
t=T \\
t=0%
\end{array}%
\right.  \label{39}
\end{equation}
coincides with the right side of the equality Eq.(\ref{38}) for the wave
function Eq.(\ref{25}), if we set $\alpha =\pi ^{-2}$. It can be expected
that the same correspondence will take place for excited states of motion of
the PU oscillator, the wave functionals of which contain polynomial
functional factors in front of the exponential in Eq.(\ref{26}).

\section{CONCLUSIONS}

The agreement between QPLA and \textquotedblleft ordinary\textquotedblright\
quantum mechanics in the local limit for the PU oscillator obtained in this
work allows us to count on the same agreement in more complex $f\left(
R^{2}\right) $ - modifications of the theory of gravity \cite{BuhLyah}. This
also serves as a justification for using QPLA in cases where the usual
canonical representation is completely absent. These include all dynamic
theories with action at a distance \cite{Fokker,WhilFey,HoylNar}.

\section{ACKNOWLEDGEMENTS}

We are thanks V.A. Franke for useful discussions.

%%%\noindent $^{\ast }$ E-mail address: alex.lukyan@rambler.ru

%%%\noindent $^{+}$ E-mail address: inna.lukyan@mail.ru

%\begin{references}

\end{document}